\documentclass[table,dvipsnames]{resonance}

\usepackage{bm}
\usepackage{verbatim}
\usepackage{tikz}
\usepackage{xcolor}

\begin{document}

\newcommand{\nwc}{\newcommand}
\nwc{\vs}{\vspace}
\nwc{\hs}{\hspace}
\nwc{\la}{\langle}
\nwc{\ra}{\rangle}
\nwc{\nn}{\nonumber}
\nwc{\Ra}{\Rightarrow}
\nwc{\wt}{\widetilde}
\nwc{\lw}{\linewidth}
\nwc{\ft}{\frametitle}
\nwc{\ben}{\begin{enumerate}}
\nwc{\een}{\end{enumerate}}
\nwc{\bit}{\begin{itemize}}
\nwc{\eit}{\end{itemize}}
\nwc{\dg}{\dagger}
\nwc{\mA}{\mathcal A}
\nwc{\mD}{\mathcal D}
\nwc{\mB}{\mathcal B}

\nwc{\Tr}[1]{\underset{#1}{\mbox{Tr}}~}
\nwc{\pd}[2]{\frac{\partial #1}{\partial #2}}
\nwc{\ppd}[2]{\frac{\partial^2 #1}{\partial #2^2}}
\nwc{\fd}[2]{\frac{\delta #1}{\delta #2}}
\nwc{\pr}[2]{K(i_{#1},\alpha_{#1}|i_{#2},\alpha_{#2})}
\nwc{\av}[1]{\left< #1\right>}

\nwc{\zprl}[3]{Phys. Rev. Lett. ~{\bf #1},~#2~(#3)}
\nwc{\zpre}[3]{Phys. Rev. E ~{\bf #1},~#2~(#3)}
\nwc{\zpra}[3]{Phys. Rev. A ~{\bf #1},~#2~(#3)}
\nwc{\zjsm}[3]{J. Stat. Mech. ~{\bf #1},~#2~(#3)}
\nwc{\zepjb}[3]{Eur. Phys. J. B ~{\bf #1},~#2~(#3)}
\nwc{\zrmp}[3]{Rev. Mod. Phys. ~{\bf #1},~#2~(#3)}
\nwc{\zepl}[3]{Europhys. Lett. ~{\bf #1},~#2~(#3)}
\nwc{\zjsp}[3]{J. Stat. Phys. ~{\bf #1},~#2~(#3)}
\nwc{\zptps}[3]{Prog. Theor. Phys. Suppl. ~{\bf #1},~#2~(#3)}
\nwc{\zpt}[3]{Physics Today ~{\bf #1},~#2~(#3)}
\nwc{\zap}[3]{Adv. Phys. ~{\bf #1},~#2~(#3)}
\nwc{\zjpcm}[3]{J. Phys. Condens. Matter ~{\bf #1},~#2~(#3)}
\nwc{\zjpa}[3]{J. Phys. A ~{\bf #1},~#2~(#3)}
\nwc{\zpjp}[3]{Pramana J. Phys. ~{\bf #1},~#2~(#3)}

\title{Stochastic energetics and thermodynamics at small scales}
\author{Sourabh Lahiri and Arun M. Jayannavar}

\maketitle
\authorIntro{\includegraphics[width=1.7cm,height=1.7cm]{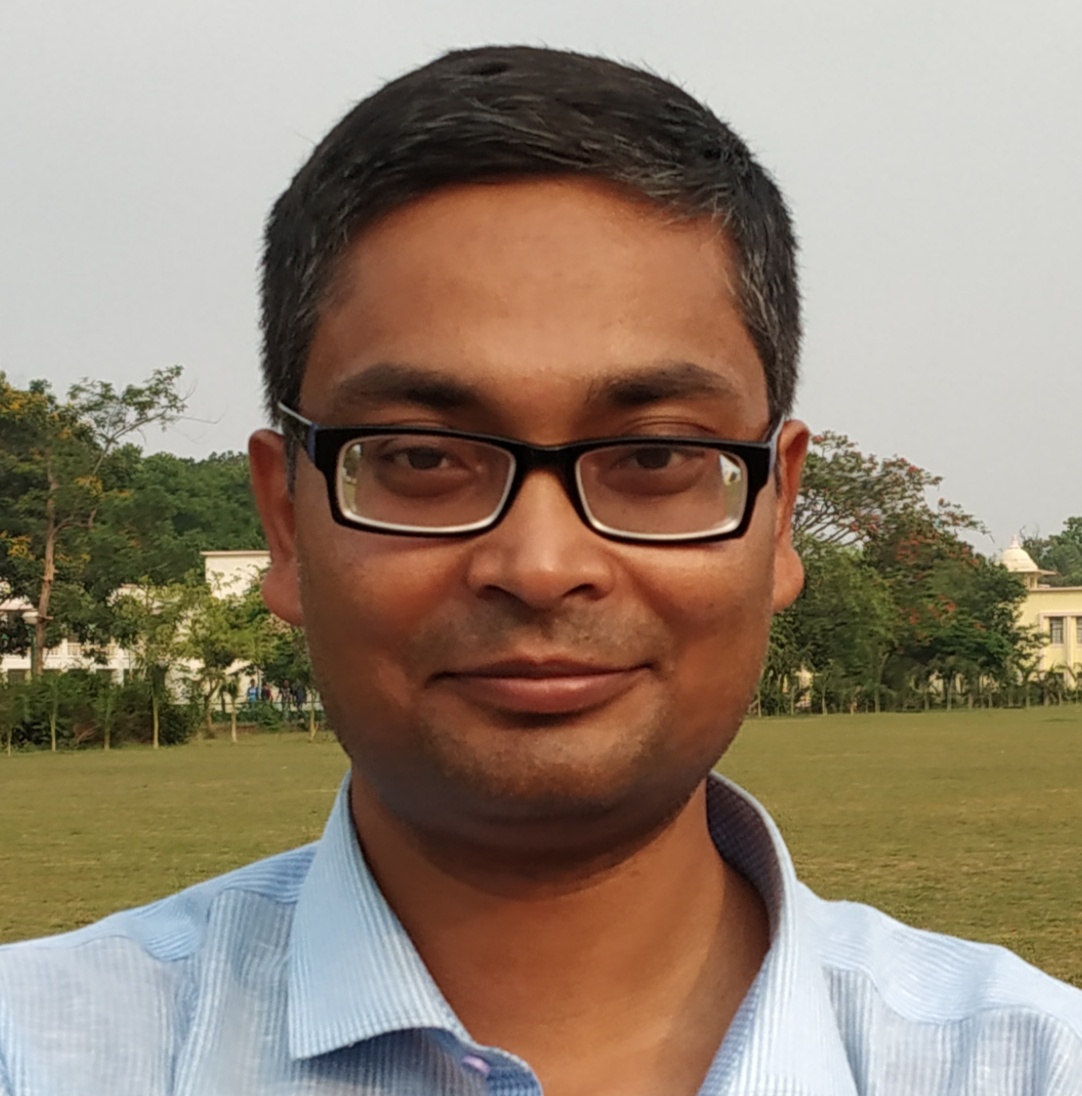}
  \includegraphics[width=1.7cm,height=1.7cm]{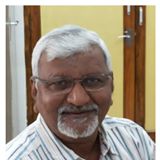}\\
  Sourabh Lahiri (left) is an Assistant Professor at Birla Institute of Technology, Mesra, and works in the field of Nonequilibrium Statistical Mechanics.\\ Arun M. Jayannavar (right) is a senior scientist at Institute of Physics, Bhubaneswar and his interests are in general Condensed Matter physics.}

\begin{abstract}
    At very small scales, thermodynamic energy exchanges like work and heat become comparable to thermal energy of the system, which leads to unusual phenomena like the transient violations of Second Law. We explore the generic characters of such systems using the framework of Stochastic Thermodynamics and provide a preliminary overview of the basic concepts. Here we have attempted to put into simple terms some actively pursued topics like the arrow of time, effect of information gain on Second Law, explanation of origin of life using Crooks theorem and the thermodynamic uncertainty relations. 
\end{abstract}

\monthyear{November 2016}
\artNature{GENERAL  ARTICLE}

\section*{Introduction}

Thermodynamics typically explains the evolution of macroscopic systems in thermal environment. The late twentieth century and thereon, however, there has been a surge in using systems that are much smaller, with the dimensions often entering the nano and meso scales (roughly 10 nm 100 nm). Given that the systems we are dealing with are not always macroscopic in nature, we are forced to extend thermodynamic definitions to incorporate such small systems in order to faciliate their study. But one important factor now comes into play: at small scales, we cannot ignore the thermal fluctuations, as one does for a macroscopic system. Thus, the extended definitions of thermodynamic quantities must necessarily possess some degree of randomness, because the experimental outcomes for the measurements of these quantities are not repeatable in the usual sense. They vary from one experiment to another, hence justifying the nomenclature \emph{Stochastic Thermodynamics}.

Nevertheless, \leftHighlight{Compare with Bohr's Correspondence Principle, which says that for large systems, the quantum system must abide by the ``usual'' classical laws} the new thermodynamics must be designed such that as the system size increases, the stochasticity automatically gets reduced and in the thermodynamic limit (number of degrees of freedom tending to infinity), one must recover the ``usual'' macroscopic thermodynamics. In the late 1990, Ken Sekimoto's intuitions made the definitions obvious \cite{sek98}. He simply transformed the Langevin equation of motion (looks like Newton's equation in presence of a damping term, with the exception of the presence of a fluctuating or random force that accounts for the random thermal kicks generated by the molecules of the surrounding medium) into a form that resembles energy conservation. Since this is nothing but the First Law of Thermodynamics, the remaining task was to algebraically manipulate the terms and rewrite the equation in such a way that the physical interpretations (which term should be called work and which one heat?) become apparent. The relevant Correspondence Principle was taken care of by the fact that the fluctuating force appearing in the Langevin equation is equally likely to assume positive and negative values, and hence vanishes on average. Thus for a large system, where we are observing the average behaviour of a large number of degrees of freedom, the stochasticity would be automatically suppressed.

 {It is to be noted that the study that will follow makes no assumption about the time of observation. The definitions appearing in Stochastic Thermodynamics are as exact for finite-time processes as they are for reversible or equilibrium processes.}

We describe briefly describe the definitions used in Stochastic Thermodynamics, as derived from a master equation.

\section{Stochastic Thermodynamics from Master Equation}

\fbox{
  \parbox{\textwidth}{
    Master equation provides the evolution of the \emph{probability distribution} of the system with time, and is typically used when the states can be represented in a discretized form. For instance, if the states at times $\{t_0,t_1,\cdots,t_N\}$ are represented by $\{x_0,x_1,\cdots,x_N\}$, both time and space being discrete, then the master equation can be written as
    \begin{align}
      \frac{dp(x_i)}{dt} &= \sum_{x_j}\left[W(x_i|x_j)p(x_j) - W(x_j|x_i)p(x_i)\right].
    \end{align}
    Here, $W(x_i|x_j)\equiv W_{ij}$ is the \emph{transition rate} from state $x_j$ to state $x_i$. The first term on the RHS is a gain term, because it provides the rate at which the system reaches state $x_i$, i.e. the rate at which the $i^{th}$ state gets populated. The second term is a loss term (hence preceded by a negative sign), because it provides the rate at which the system leaves the $i^{th}$ state, i.e. the rate at which this state gets depopulated.

    As a side note, if one wishes to keep the states continuous but the time discrete, the RHS of the master equation would simply have its summation replaced bt an integration over $x_j$.
    }
  }

  We write the master equation as
  \begin{align}
     \frac{dp(x_i)}{dt} &= \sum_{x_j}\left[W(x_i|x_j)p(x_j) - W(x_j|x_i)p(x_i)\right].
  \end{align}
  Note that the states $x_i$ can represent any state variable and not just position. For instance, we may consider a quantum particle trapped in a box, and represent the energy states using this notation.

  Gavin Crooks introduced a very useful and intuitive way of looking at the evolution \cite{cro99_pre}. In each time step, the system energy changes, and this energy change occurs at two levels:
  \begin{enumerate}
  \item External parameter remains constant, but state changes by gaining (loosing) energy from (to) the environment. This was termed as the \textbf{heat step}, because the energy change in this step can only occur due to exchange of heat energy between the system and its environment.

  \item The state remains same, but external parameter changes. This must be the \textbf{work step}, where the work done on the system (by the external parameter) accounts for the change in its energy.
  \end{enumerate}
  For visualizing the problem better, consider a quantum particle in a one-dimensional box whose walls are conducting and in contact with an external heat bath. The width of the box is $\lambda$ (say), and the energy levels are given by
  \[
    E_n = \frac{n^2\pi^2\hbar^2}{2m\lambda^2}, \hspace{1cm} n=1,2,3,\cdots
  \]
  Suppose that the particle is initially in the ground state. It then jumps to its first excited state $n=2$, with the width of the well (externally controlled parameter) fixed at the value $\lambda$. Since the external force is not doing any work, and yet there is a change in the energy of the particle, the inevitable conclusion is that this energy must be coming from the surrounding heat bath. Thus, this step is termed as the \emph{heat step}. The absorbed heat is given by $\delta Q = E_2(\lambda) - E_1(\lambda)$.

  Next, suppose the external parameter changes from $\lambda$ to $\lambda'$, but the particle remains in the state $n=2$. Once again, there is a change in energy, this time because the external force is doing work on the particle. This step is thus the \emph{work step}. The work done on the particle in this step is given by $\delta W = E_2(\lambda')-E_2(\lambda)$.

  If the observer only observes the net effect of these two steps ($\lambda\to\lambda'$ and $n=1\to n=2$), he will find that the net energy change is $\delta E = E_2(\lambda')-E_1(\lambda)$. It is trivial to check that $\delta E = \delta Q + \delta W$, which is the \emph{First Law} for the time step under consideration. Since this holds for each step, it also holds for an entire trajectory in state space, because it is made up of a large number of such small steps.  The state need not necessarily be described by energy; it can be any observable property of the system (see figure \ref{fig:traj}, where three typical trajectories have been shown and $x$ denotes the state describing the system). Thus, the First Law becomes
  \begin{align}
    \Delta E &= Q + W,
  \end{align}
  where $\Delta E = \sum_{i=1}^N \delta E_i$, $Q = \sum_{i=1}^N \delta q_i$ and $W= \sum_{i=1}^N \delta W_i$, where the total time period of observation has been divided into $N$ small steps.
  
  \begin{figure}
      \centering
      
  \begin{tikzpicture}[scale=1.3,every node/.style={scale=1.3}]
    
  \draw[very thick,->] (0,0) -- (7,0);
  \draw[very thick,->] (0,0) -- (0,2.5);
  \draw[very thick,red,->] (0,1) .. controls (1,1.5) .. (4,1) .. controls (5,0.7) .. (6,1);
\draw[very thick,blue,->] (0,1) .. controls (1.5,0.4).. (3,1) .. controls (4.5,1.6).. (6,1.4);
\draw[very thick,brown,->] (0,1) .. controls (1.2,1.2) .. (2,0.8) .. controls (2.5,0.4) .. (4,0.8).. controls (5,1.2).. (6,0.5);
  \node[] at (6,-0.3) {$t$};
  \node[] at (-0.3,2) {$x$};
  \node[] at (-0.3,1) {$x_0$};
  \end{tikzpicture}
  \caption{Some typical trajectories that are produced during a process, where the initial point has been fixed at $x_0$.}
  \label{fig:traj}
  \end{figure}
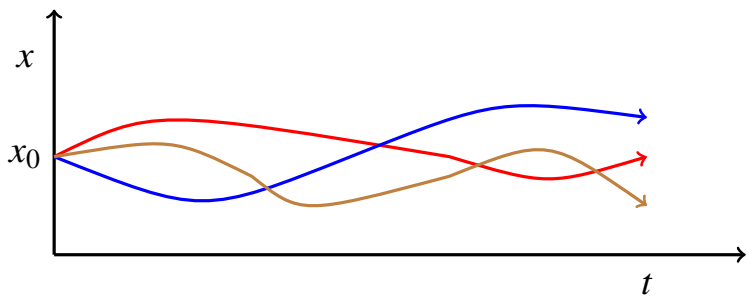

\section{Second Law and the Arrow of Time}

If we observe a process in which shards of glass assemble themselves together and construct a beautiful piece of vase, we somehow ``know'' that the process is not real, it must be a movie that is being played backward. \rightHighlight{The entropy, roughly speaking, is a measure of the intrinsic randomness of a system. A drop of ink concentrated at a given location in a glass of water has far less entropy as compared to the entropy after the ink has diffused throughout the body of water} This knowledge comes from common sense, where often we find that the reversed sequence of several processes are \emph{never} encountered. But why not, when there is no reason to rule them out using arguments of conservation of energy? Here steps in the Second Law, and dictates that even though a process is not in contradiction with energy conservation, it is not allowed if the total \emph{entropy} gets reduced during the process. The process is not outright unfeasible, but is extremely improbable, so much so that one might need to wait for a time span that far exceeds the estimated lifetime of the universe to record even a single such event.
The arrow for quasi-isolated classical and quantum systems has  been investigated in \cite{yuk03_pla,yuk12_pla} by Yukolav. He explains how an expansion in phase space volume can be a reliable measure measure of irreversibility (especially if dynamical systems are taken into account) \cite{yuk03_pla}. The relation between loss of information in a quantum system and arrow of time has been discussed in \cite{yuk12_pla}.

Such intuitive ``knowledge'' that one gains from daily experiences can actually be quantified, as was observed by Jarzynski in \cite{jar10_arcmp}. This provides the notion of the arrow of time, which means that flow of time is always along a given (forward) direction, and a reversal in its direction is highly unlikely. The forward direction is one along which the total entropy increases, while a reversal in its direction would imply that the process entails a reduction of entropy.

The arrow, however, begins to get fuzzy as one focusses on systems which are of mesoscopic dimensions. This is the regime where the energy exchanged with surroundings are of the order of thermal energy of the system. In this regime, there is an appreciable probability that one might actually observe a atypical (unexpected) sequence of events,  {often referred to as the \emph{transient violations of Second Law}}. It was beautifully quantified by Jarzynski in the form of a \emph{guessing game}, where by watching a movie, one has to guess whether it is a real process, or it is a movie that is being played backward. He provides the exact probability for making a wrong guess, thereby exhibiting a process that unfolds in a direction opposite to the arrow of time.

\section{Fluctuation Theorems and the effect of information}

Now let us visualize the diffusion of gas particles in terms of a large number of particles confined in a (classical) box. If the particles are more or less evenly distributed throughout the box, we will never observe the particles to gather in one half of the box. However, if there are two particles in the box, one in each half, then there is an appreciable probability (25\% to be precise) of observing both the particles in one of the halves at some later time. So we see that as the number of degrees of freedom becomes very small as compared to a macroscopic system, the arrow of time gets blurred. The Fluctuation Theorems are a set of relations that help in quantifying the extent of this fuzziness.

Maxwell had proposed his famous ``demon'' in 1867, which is an intelligent being that is able to manoeuvre individual molecules of a gas. A box containing a gas is divided into two chambers, and a demon is in charge of a trap door that separates the chambers. It allows fast moving particles (those that have speeds greater than the mean speed) to go only from left to right, while the slow particles are allowed to move only from right to left. This eventuated in a situation where the right chamber got heated up, and the left chamber got cooled down, thus leading to the violation of the Second Law. In 1929, Leo Szilard put forward a simpler version of this demon in the form of an engine, famously know as the \emph{Szilard Engine}, working with a single molecule of gas. He showed that it is possible for the engine to work in a cycle, and convert the heat energy of the surrounding thermal bath into work, again in apparent contradiction of the Second Law. However, after several decades of engaging debates, it was finally realized by Landauer that the problem lies in one of the steps in the working cycle of the engine, namely where the external controller measures the position of the particle. The full cycle is not complete as long as this information acquired at the intermediate step is not erased. Conversely, if the information is not erased, then the form of the Second Law must be modified to  account for this fact. He showed that this erasure process involves the dissipation of $k_BT\ln 2$ amount of energy \emph{per bit} of information, which removes the  paradox of violation of Second Law. Moreover, it is to be emphasized that the information erasure is a physical process, as the inforation is coded in a physical system (``information is physical''). \emph{These are some examples that have paved the way for the development of Stochastic Thermodynamics. Conversely, the formulation of Stochastic Thermodynamics has helped us in gaining better insights into these events, and has established the Landauer erasure principle on a firm ground.}

What, then, is the modified Second Law in presence of information? The result is given by
\begin{align}
    \av W \ge \Delta F-k_BT \av I.
    \label{MSL}
\end{align}
Here, $\Delta F$ is the change in the equilibrium free energy during the process,  {which equals the amount of \emph{reversible} work that can be done on the system. In absence of information, the second term on the right hand side vanishes, and it yields the Maximum Work Theorem, which says that the extracted work is maximum (or the input work is a minimum) for a reversible process}. $\av I$ is the mean information gained about the \emph{actual state} $x$ of the system, by examining the \emph{measured state} $m$, using the fact that the actual and the observed states are correlated. Of course, this correlation is strong if the measurement is accurate, otherwise it is weak.  {The quantity $I$ is referred to as \emph{mutual information} in the literature, since it quantifies the mutual dependence of $x$ and $m$ on each other.}

\fbox{
\parbox{\textwidth}{
The average mutual information between any two variables, say $y$ and $z$, is defined as a \emph{distance} between the probability distributions $p(y,z)$ (joint probability of the two variables) and $p(y)p(z)$ (product of the marginal probabilities of each variable). The ``distance'' is referred to as the Kullback-Leibler divergence, defined as
\begin{align}
    D[p(y,z)||p(y)p(z)] &\equiv \int dy dz ~p(y,z)\ln\left[\frac{p(y,z)}{p(y)p(z)}\right] \nn\\
    &= \int dy dz ~p(y,z)\ln\left[\frac{p(y|z)}{p(y)}\right]\nn\\
    &= \left< \ln\left[\frac{p(y|z)}{p(y)}\right]\right>.
\end{align}
The second line follows from the first one on application of Bayes' Theorem: $p(y,z)=p(y|z)p(z)$.
}
}

\leftHighlight{
If the measuring device is infinitely inaccurate, i.e., the outcome $m$ has no correlation with the correct value $x$, then the mutual information vanishes, as it should
} If the error leading to wrongly observing $x$ as $m$ is given by the conditional probability $p(m|x)$, then the information gained, or the \emph{mutual information} as it is known in the literature, is defined as
\begin{align}
    I \equiv \ln\left[\frac{p(m|x)}{p(m)}\right].
\end{align}
The denominator is the marginal probability density for obtaining the value $m$. 

Eq.\eqref{MSL} \leftHighlight{The new result reveals that information is a physical quantity and can change the effective free energy of the system} immediately tells us something that is counter-intuitive and seems to violate the ``usual'' Second Law, namely that the work done on the system can be less than the difference in free energy. Simply put, if the process is cyclic (so that there is no change in free energy), work can in fact be \emph{extracted} from the system, and provides a way to harness enegry from the random thermal motion of the particles of the medium. However, keeping in mind that information is now on a similar footing as other thermodynamic variables, this should come as less of a surprise. 

Information thermodynamics has steadily established itself as one of the most useful and indispensable aspects of thermodynamics. The so-called \emph{information engines}, that work on the principle of delivering work at the expense of information, have been analyzed both theoretically and experimentally. 

\fbox{
\parbox{\textwidth}{
 The Second Law connects mean work to free energy change through an inequality: $\av W\ge \Delta F$.
  Here, $W$ is the work done \emph{on} the system, and $\Delta F$ is the change in the \emph{equilibrium} free energy in the process. The Jarzynski equality, which is one of the Fluctuation Theorems, relates these quantities by means of an equality, which makes it a much stronger statement:
  \begin{align}
      \av{e^{-\beta W}} = e^{-\beta \Delta F}.
      \label{JE}
  \end{align}
  Now, using the convexity of the exponential function, one obtains the Second Law inequality as a corollary (known as Jensen's inequality, which in the particular case of an exponential function states that $\av{e^{-y}}\ge e^{-\av y}$, where $y$ is the random variable in question).
  In presence of information, this relation changes to
  \begin{align}
            \av{e^{-\beta W-I}} = e^{-\beta \Delta F}.
  \end{align}
  Again applying the Jensen's inequality, we get
  \begin{align}
      \av W \ge \Delta F - k_B T \av I,
  \end{align}
  which is eq. \eqref{MSL}.
  In a cyclic process, $\Delta F=0$, but from Eq. \eqref{MSL}, it can be seen that work can still be extracted during the process in presence of information. Thus information acts as a resource.
  }
  }

  \section{Importance of Stochastic Thermodynamics and Fluctuation Theorems}

  One significance of these relations is the elucidation of Second Law. 
  They are stronger relations than the Second Law, being exact equalities in contrast to the latter. It also has provided impetus to several areas of research. One of the most prominent areas is probably the study of \emph{Origin of life} \cite{jeremy}, where the author explains how the fluctuation relation for entropy can be used to understand the phenomenon of self-replication. 
  \leftHighlight{The linear response theory works well when the perturbing force is not too large, so that the response to this perturbation can be assumed to be linear in the strength of the perturbation itself. For instance, a mechanical force on a Brownian particle produces a change in its velocity. The latter is often related linearly to the applied force.} Another area that is based on stochastic thermodynamics is the study of \emph{information engines}, which can convert information into mechanical work \cite{sei12_jpa,gov18_prl}. 
  It might be one of the directions in which future research on nanomachines can focus on. Then there are \emph{thermodynamic uncertainty relations} \cite{sei15_prl}, where the authors show that such a relation exists between an observable and the free energy cost necessary to give rise to such an observable. Last but not the least, the fluctuation relations have been used in finding symmetry relations among \emph{nonlinear response coefficients}, thus exhibiting the superiority of these relations over the well-established linear response theory \cite{gas04_jsm} that can provide symmetry relations among the linear response coefficients only.


  \section{Relation to origin of life}

  It was known that entropy change is a measure of irreversibility of a process. This measure was quantified by Seifert \cite{sei05_prl}, when he showed that the logarithm of the ratio between the probability of a forward trajectory and that of its corresponding reversed trajectory, gives the entropy produced \emph{along the forward trajectory}. In short, the mathematical statement for the average change in total entropy is
  \begin{align}
    \Delta S_{tot} \equiv \av{\ln\frac{Prob(\mbox{Forward Trajectory})}{Prob(\mbox{Reverse Trajectory})}}.
  \end{align}
   {This physically means that the magnitude of change in total entropy determines how reversible a process is. It was realized by Jeremy England \cite{jeremy} that this irreversibility must be manifested in processes such as self-replication and evolution, so that the concepts must fall, to some extent, within the purview of Statistical Physics. }
  It readily gave the Second Law inequality as a corollary: $\Delta S_{tot}\ge 0$ (i.e., entropy of the universe never decreases with time).
 
  This definition was used in \cite{jeremy} to derive the corresponding ratio between two \emph{macroscopic states}, say ${\bf I}$ and ${\bf II}$ (see fig. \ref{fig:traj}). There may be numerous microstates associated with each such macrostate, but when he integrated over the transitions between all such microstates (consistent with the constraint of being in a particular macrostate), he arrived at an inequality that ``looks'' like the Second Law:
  \begin{align}
    \beta\av{\Delta Q}_{{\bf I}\to{\bf II}} + \ln\frac{P({\bf II}\to{\bf I})}{P({\bf I}\to{\bf II})} + \Delta S_{int}  \ge 0.
  \end{align}
   {The first and the third terms are the heat dissipated and entropy change while going from \emph{macro}state {\bf I}  to \emph{macro}state {\bf II}. The second term is the logarithm of the ratio between probability of going from state {\bf I} to {\bf II} and its reverse, and this term differentiates the inequality from the normal Second Law.}

  \begin{figure}
      \centering
      
  \begin{tikzpicture}[scale=1.3,every node/.style={scale=1.3}]
    \draw[fill=yellow] (1,1) ellipse (0.5cm and 0.7cm);
    \draw[fill=yellow] (6,1) ellipse (0.3cm and 0.5cm);
  \draw[very thick,->] (0,0) -- (7,0);
  \draw[very thick,->] (0,0) -- (0,2.5);
  \draw[very thick,red,->] (1,1) to [out=70,in=200] (6,1);
  \draw[very thick,blue,<-] (1,0.8) to [out=70,in=200] (6,0.8);
  \node[] at (4,1.6) {\color{red} Forward};
  \node[] at (3,0.8) {\color{blue} Reverse};
  \node[] at (6,-0.3) {$t$};
  \node[] at (-0.3,2) {$x$};
  \node[] at (1,2) {\bf I};
  \node[] at (6,1.8) {\bf II};
  \end{tikzpicture}
  
  \caption{Schematic diagram of a forward and reverse trajectory in a one-dimensional state space. A single forward trajectory and its reverse are shown, connecting two macrostates {\bf I} and {\bf II}.}
  \label{fig:traj}
  \end{figure}
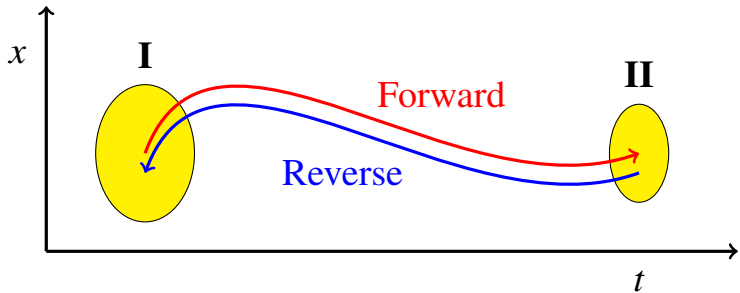

  In \cite{jeremy}, the author was successful in deriving the amount of heat (obtained from the macroscopic fluctuation theorem) dissipated by a self-replicating RNA molecule. He also showed, using similar treatment, that a DNA molecule is thermodynamically far more stable (against self-replication) than an RNA molecule.

  The author advances further to the more complex example of bacerial cell division, and was able to provide a rough lower bound for the amount of heat dissipated in the process.

  In a more recent work \cite{eng18_arxiv}, the authors quantify the conducive physical conditions needed for efficient self-replication. He used a typical system consisting of a large number of chemical species undergoing various reactions, that involved autocatalytic cycles (one of the reactants catalyzes its own production), coupled to lossy side-reactions (other reactions that interfered with the autocatalytic cycles to use up some of the resources).

  Several protocols were used and in each case, conditions that led to self-replication were explored. The authors finally note that the assumption that the resources were abundant throughout the time-scale of the experiment might introduce deviations from realistic results, apart from other factors that might affect the results of this work in real systems.
  
  \leftHighlight{Think about a person pushing a block up an inclined plane. If this is done very fast, the invested energy will be much higher than the potential energy gained by the block, due to dissipation caused by frictional forces}In another seminal paper, he puts forward the idea of \emph{dissipative adapatation}, where a certain configuration of a system of particles becomes efficient in absorbing and dissipating work, and thereby changing the configuration. The more irreversible a process is, the more work it must dissipate under similar external conditions. Considering that life processes are typically far from equilibrium, all living organisms must absorb and subsequently dissipate large amounts of work (as compared to a process that is nearly reversible). He goes on to show that if we assume some microstate $i$ to be equally accessible kinetically by a large number of microstates, 
  then the most likely outcome of a measurement within the latter domain 
  would be the state that maximizes dissipation of work in the process that takes the system from set $i$ to the set of the final microstates.
  
  \section{The Thermodynamic Uncertainty Relations}

  Neils Bohr had once mentioned that an uncertainty relation between energy and temperature must exist in statistical physics, in a manner similar to that in quantum mechanics. In 1988, Sch\"ogl developed the uncertainty relations between thermodynamically conjugate variables. As an example, one can show that the product of a fluctuation in temperature and a fluctuation of the energy of the system is never less than unity. 

  The concept has undergone a reinvention in recent years, and has attracted a lot of attention. Broadly, the statement is as follows \cite{sei15_prl}:
  \emph{There exists a fundamental relation between the uncertainty in measuring an observable and the free energy cost required to maintain the underlying process which gives rise to that observable.}
 {This has been used to understand and analyze the reliability of emergence of self-organization and replication in living systems.}
  Mathematically, for any current $j$ (heat current, for instance) it can be written as
  \begin{align}
    \frac{\mbox{Var}[j]}{\av{j}^2} &= \frac{2k_B}{\Sigma},
    \label{TUR}
  \end{align}
  where $\Sigma$ is the total entropy production, and $k_B$ is the Boltzmann constant. $\mbox{Var}[j]$ and $\av{j}$ are the variance and mean of the current, respectively. If $\sigma$ is the entropy production \emph{rate}, then $\Sigma=\sigma t$. %

  The above relation implies that there exists a trade-off between the precision of measurement and the associated energy dissipated during the process. If a very precise value of the observable is required, the dissipation cost would be much higher.

  As an example, in a biased random walk, the current is simply the ratio between net number of steps towards right, and time taken. If the entropy production rate is small during the process, it is not possible to obtain a very precise value of the final position of the particle on the lattice.
  
   {In a recent paper \cite{has19_arxiv}, the authors have derived a weaker form of the uncertainty relations from the Fluctuation Theorem for total entropy. The study has been extended to feedback-controlled protocols in \cite{has19a_arxiv}.
  }
  
  \section{Stochastic Thermodynamics for quantum particle}
  
    The stochastic thermodynamics of quantum particles has gained a lot of attention over the last decade. It was realized that work is not an observable \cite{han07_pre}, in the sense that there is no operator whose eigenvalue can provide the value of work. Nevertheless, it was deftly defined by Hanggi and coworkers \cite{han11_rmp}. They pointed out that although for the system of interest the definition of work is not obvious, one can define it by measuring the difference in energy of the combined system consisting of the system and the heat bath. This is because the gain in energy of the composite system can only arise from work being done on the system of interest. They also derived the Jarzynski Equality for the quantum particle, and showed that the form is same as that in the case of a classical particle. Nevertheless, the field is under intense study in order to gain further understanding of the thermodynamics.
  
\section*{Acknowledgement}

One of us (AMJ) thanks DST, India for financial support (through J. C. Bose National Fellowship). SL thanks DST-SERB (India) (project sanction no. ECR/2017/002607) for financial support.

\bibliographystyle{unsrtnat}

\end{document}